# Modes trimming and clustering in a weakly perturbed high-Q whispering gallery microresonator

Botao Fu,[1,7] Renhong Gao,[1,6] Jintian Lin,[1,6] Ni Yao,[2,†] Haisu Zhang,[3] Min Wang,[3] Lingling Qiao,[1,6] Wei Fang,[5] and Ya Cheng[1,3,4,6,8,9,10,11,*]

[1] State Key Laboratory of High Field Laser Physics and CAS Center for Excellence in Ultra-Intense Laser Science, Shanghai Institute of Optics and Fine Mechanics (SIOM), Chinese Academy of Sciences (CAS), Shanghai 201800, China

[2] Research Center for Humanoid Sensing, Zhejiang Lab, Hangzhou 311100, China

[3] The Extreme Optoelectromechanics Laboratory (XXL), School of Physics and Electronic Science, East China Normal University, Shanghai 200241, China

[4] State Key Laboratory of Precision Spectroscopy, East China Normal University, Shanghai 200062, China

[5] Interdisciplinary Center for Quantum Information, State Key Laboratory of Modern Optical Instrumentation, College of Optical Science and Engineering, Zhejiang University, Hangzhou 310027, China

[6] Center of Materials Science and Optoelectronics Engineering, University of Chinese Academy of Sciences, Beijing 100049, China

[7] School of Physical Science and Technology, Shanghai Tech University, Shanghai 200031, China

[8] Collaborative Innovation Center of Extreme Optics, Shanxi University, Taiyuan 030006, China

[9] Collaborative Innovation Center of Light Manipulations and Applications, Shandong Normal University, Jinan 250358, China

[10] Shanghai Research Center for Quantum Sciences, Shanghai 201315, China

[11] Hefei National Laboratory, Hefei 230088, China

[†] Electronic address: yaoni14@zju.edu.cn
[*] Electronic address: ya.cheng@siom.ac.cn

**Abstract**

In general, a high-Q microresonator can accommodate abundant whispering gallery modes (WGMs) with the mode number increasing with the dimensional sizes of the microresonator. Removing the unnecessary modes while reorganizing the remaining modes is of vital importance, which, however, has been proved challenging and usually results in a tradeoff with the Q of the microresonator. Here, we reveal an effective and controllable mode trimming and clustering mechanism underlying the generation of polygon and star modes in weakly perturbed tapered fiber-coupled lithium niobate whispering gallery microresonators. Experimentally, various polygon and star modes are observed in sequence within a single microresonator by tuning the excitation wavelength or varying the coupling position between a tapered fiber and the circular microresonator, which can be well reproduced with our theoretical model. The finding offers a ubiquitous solution for a broad range of applications requiring elaborate selection and organization of the high-Q WGMs.

Optical dielectric microresonators, which enhance light-matter interaction by storing optical energy in a small mode volume with ultra-low loss [1-4], are now highly desirable not only to fundamental physics such as chaos-assisted tunnelling [5, 6] and cavity quantum electrodynamics [7, 8] but also to practical applications from integrated laser sources [9-12], to sensors [13-15] and modulators [16, 17]. Among the diverse candidates, circular-shaped whispering gallery mode (WGM) microresonators have been intensively studied due to their nearly perfect light confinement over a long period. The isotropic characteristic in the microdisk plane is unfavourable for either intensive interaction of the light field with the coupled device or efficient emission collection, which is the reason behind the significant efforts spent on the deformed WGM microdisks [18-21]. Unfortunately, the microresonator deformation usually leads to the reduction of the mode Q factor and in turn the light-matter interaction strength. Recently, the generation of high-Q polygon modes in tapered fiber-coupled circular microdisk resonators with weak perturbation has been reported [22, 23], whereas the essential physics behind the phenomena is yet to be explored and clarified.

Here, we reveal an effective and controllable mechanism to precisely conduct the mode trimming and clustering in a weakly perturbed tapered fiber-coupled lithium niobate WGM microresonator. Meanwhile, complete polygon and star modes sequences are observed within the single microresonator, which are well reproduced with our theoretical model.

To facilitate observation of mode patterns, an $Er^{3+}$-doped z-cut lithium niobate circular microdisk resonator allowing visualization of the modes is utilized taking advantage of two-photon fluorescence excitation of the $Er^{3+}$ ions. The circular microdisk resonator was fabricated using photolithography-assisted chemo-mechanical etching, and further details on the fabrication procedures can be found in Ref. [24-26]. The inset of Fig. 1(a) shows the scanning-electron-microscope (SEM) image of the fabricated microdisk resonator with a thickness of 700 nm and a diameter of 83.44 μm.

The experimental setup for investigating the polygon and star modes formation is illustrated in Fig. 1(a), in which a narrow-linewidth tunable diode laser with a wavelength range from 960 nm to 980 nm (Model: TLB-6719-P-D, New Focus, Inc.) is used as the pump light source. The pump light is coupled to the microdisk via the tapered fiber with a diameter of 2 μm [27], and the polarization state can be controlled to be transverse-magnetically polarized by an in-line polarization controller (PC). The input pump power can be adjusted and monitored by a variable optical attenuator (VOA) and a power meter combined with a 10/90 beam splitter. The relative position between the tapered fiber and microdisk resonator can be monitored and adjusted by an optical microscope system and a high-resolution positioning stage of 5 nm resolution. A top-view optical microscope with a 20× objective lens with a numerical aperture of 0.42, a visible charge-coupled device (CCD), and a short-pass filter (Model: FES800, Thorlabs, Inc.) are used to capture the mode patterns distributed on the plane of the microdisk. Remarkably, the captured up-conversion fluorescence patterns directly reflect the spatial distribution characterisitics of the pump mode [23].

Experimentally, to excite the polygon and star modes with high stability, the tapered fiber must be placed carefully on the top surface being in contact with the circular microdisk resonator. When the distance between the tapered fiber and the microdisk center is tuned to 38.5 μm, the triangle mode (ground state), the dual-localization triangle mode (first excited state) [28] and the star mode can be excited at varied wavelengths of 972.47 nm, 972.56 nm and 972.83 nm, as demonstrated in Figs. 1(b), 1(c) and 1(d), respectively. When further increases the relative distance between the tapered fiber and the microdisk center to 39.8 μm, the square mode, pentagon mode, hexagonal mode, heptagon mode, and octagon mode can be successively excited at increasing wavelengths of 972.98 nm, 974.48 nm, 975.65 nm, 978.82 nm, and 979.86 nm, as shown in Figs. 1(e)-1(i), respectively. Thus, various polygon and star modes are sequentially observed in the single microdisk by synergetically manipulating the pump wavelength and the tapered fiber position.

To shed light on the mode trimming and clustering mechanism, we first reveal the mode distribution characteristics of the whispering gallery microresonator. For the thin circular microdisk with a diameter of $R$, the general form of the eigenmodes in cylindrical coordinates ($r$, $\theta$) can be expressed as follows [29],

$$\psi(r,\theta) = \begin{cases} J_m(nkr)e^{im\theta} & r \leq R \\ H_m^{(1)}(kr)e^{im\theta} & r \geq R \end{cases} \quad (1)$$

where $m$ ($m=0,1\ldots$) is an integer that specifies the azimuthal quantum number related to orbital momentum, $J_m(x)$ ($H_m^{(1)}(x)$) refers to the $m$th order Bessel function (the $m$th Hankel function of the first kind), $n$ and $k$ denote the effective index of the two-dimensional thin film and vacuum wave vector, respectively. By imposing the boundary condition of the electromagnetic field, the quantization condition can be expressed in the following form,

$$S_m(nkR) = \frac{n}{\nu}\frac{J'_m}{J_m}(nkR) - \frac{H_m^{(1)'}}{H_m^{(1)}}(kR) = 0 \tag{2}$$

here $J'_m(x)$ and $H_m^{(1)'}(x)$ represent the first order derivatives of the $m$th order Bessel function and the $m$th order Hankel function of the first kind, respectively, and $\nu$ is 1 ($n^2$) for transverse magnetic (transverse electric) polarization. The quantization condition can determine the quantum numbers of the microcavity modes, and the eigenvalues obtained in the microresonator can be defined as [30],

$$nk_{p,m}R = x_{p,m} + \delta x_{p,m} \tag{3}$$

with $\delta x_{p,m} \ll x_{p,m}$, the correction term $\delta x_{p,m}$ introduced by the boundary conditions depends merely on the geometry characteristics of mode chord angle [28], $p$ ($p$=1,2…) is an integer that describes the radial quantum number, and $x_{p,m}$ is the $p$th root of $m$th order Bessel function. Physically speaking, the wave vectors $k_{p,m}$ of quasistationary states depend on radial quantum number $p$ and azimuthal quantum number $m$. The eigenvalue interval of arbitrary WGMs can be described as,

$$x_{p-Q,m+P} - x_{p,m} = \frac{P}{\sin\overline{\Theta}}\left(\overline{\Theta} - \frac{Q\pi}{P}\right) + O(x^{-2}) \tag{4}$$

where $P$ and $Q$ denote the changes of the azimuthal and radial quantum numbers, respectively, $\Theta$ describes the mode chord angle with a relation of $\cos\Theta = m/x$, and $\overline{\Theta}$ is the angle between $\overline{\Theta}_{p,m}$ and $\overline{\Theta}_{p-Q,m+P}$. The cosine of chord angle is dominated by the azimuthal quantum number $m$ when the eigenvalue is restricted to a specific range. According to Eq. (4), those modes with eigenvalues of $x_{p,m}$ and $x_{p-Q,m+P}$ are degenerate when the chord angle $\overline{\Theta}$ of the mode is close to $\frac{Q\pi}{P}$. A more detailed discussion of the eigenvalue characteristics can be found in Suppl. Mater. S1. For the quasi-degenerate mode sequence $\left\{x_{p,m}^{(P,Q;M)} = x_{p_0-MQ,m_0+MP} | M = 0, \pm 1, \ldots \ldots \pm M_m\right\}$, the eigenvalue difference can be further represented as,

$$x_M - x_0 = -\frac{P^2}{2x_0 \sin^2(Q\pi/P)} M(M + \Delta_m) + O(x^{-2}) \tag{5}$$

with $x_M = x_{p,m}^{(P,Q;M)}$, and $M$ is the sequence number of quasi-degenerate modes. The $\Delta_m$ describes the mismatch angle between $\overline{\Theta}_0$ and $\frac{Q\pi}{P}$ for the classical trajectory and varies periodically with angular quantum number, resulting in accidental degeneracy.

The quasi-degenerate modes with distinct geometric features widely exist in the microresonator, as demonstrated by the black dots in Fig. 2. The eigenvalues are distributed at different parabolic curves fitted by Eq. (5) centered with $\cos\Theta$ equals to 0.5, 0.707, 0.809, 0.866, 0.901, and 0.924, where $P$ equals to 3, 4, 5, 6, 7 and 8, respectively. And the interval between the center points of the curve is $\Theta/\sin\Theta$, indicating an excellent consistency with the classical polygon modes. In contrast, the interval between the parabolic curves is reduced by half at the position of $\cos\Theta|_{2\pi/7} = 0.623$, which is consistent with the characteristics of the seven-star mode. As indicated by the different mode distribution features, the phase space of modes can be divided into two categories, namely, the regions of non-uniform distribution labelled by dark color near the center of high degeneracy and that of uniform distribution marked by light color at the boundary of the non-uniform regions. In the latter regions of the two featuring the uniform distribution, the polygon and star modes can not be easily formed due to the low probability of mode degeneracy. In the former regions of non-uniform distribution, the neighbouring WGMs with the fixed $P$-difference of the angular quantum number allow the mode recombination, forming the polygon and star modes with $P$-fold symmetry.

The influence caused by the tapered fiber can be regarded as a refractive index modulation $\Delta$ generated at the position $r = R(1 - \lambda f(\theta))$, where $\lambda f(\theta)$ reflects the geometric characteristics of the modulation. Imposing the new boundary caused by tapered fiber of the electromagnetic field, the quantization condition can be expressed in the following form,

$$\sum_m a_m S_m(x) \cos(m\theta) = \lambda x f(\theta) \sum_m a_m \frac{\partial}{\partial x} S_m^{(\Delta)}(x) \cos(m\theta) \qquad (6)$$

where $S_m^{(\Delta)}(x)$ and $S_m(x)$ denote the quantization condition with and without perturbation induced by the tapered fiber, respectively. The rates of changes of quantization conditions $\frac{\partial}{\partial x} S_m^{(\Delta)}(x)$, which are proportional to the difference in refractive index inside and outside the boundary, reflect mode constraint capacity (see, Suppl. Mater. Eqs. S1-15). Obviously, the rotational symmetry of the microcavity is broken due to the coupling of the tapered fiber, and a more detailed discussion can be found in Suppl. Mater. S3.

The zero-order equation of the quantization condition in degenerate subspace can be obtained by multiplying $\{\cos m'\theta \,|\, m' \in P_0)\,\}$ and by summing both sides over all $m'$, we get,

$$S_{m\in P_0}(x_0 + \lambda x_1)a_{m\in P_0} = -\frac{2\lambda\Delta}{n^2-1}x_0\frac{\partial}{\partial x}S_{m\in P_0}(x)\sum_{m'\in P_0}A_{m',m}a_{m'} \quad (7)$$

here $x_0$ is the mode eigenvalue of microresonator without perturbation, $P_0 = \sum P_0^{(\Theta)}$ is the degenerate subspace around the eigenvalue, $\frac{-2\Delta}{n^2-1}$ is the ratio of mode constraint capacity with and without the tapered fiber, and the metrix $A_{m',m} \cong \frac{1}{2\pi}\int_0^{2\pi}f(\theta)\cos((m'-m)\theta)d\theta$ denotes the degree of mode coupling in the degenerate subspace $P_0$. In particular, modes with different chord angles hardly interact even under the influence of the tapered fiber, then the relation of $A_{m'\in P_0^{(\Theta_1)},m\in P_0^{(\Theta_2)}} \ll A_{m'\in P_0^{(\Theta_1)},m\in P_0^{(\Theta_1)}}$ is acquired. Thus, degenerate modes with different families can be discussed independently, and combined with Eq. (5), each family described by the quasi-degenerate sequence $\{x_{p,m}^{(P,Q;M)}\}$ will be recombined,

$$\sum_M\left\{[-\lambda x_1 + f_{P,Q}M(M+\Delta_m)]\delta_{M,M'} - \frac{2\lambda\Delta e^{-i\gamma\,sign(M-M')}x_0}{n^2-1}A_{M,M'}\right\}a_{M'} = 0 \quad (8)$$

where $a_M = \langle p_0 - MQ, m_0 + MP|\psi_{new}\rangle$ represents the proportion of WGMs in new recombination states, $\lambda x_1$ is the eigenvalue correction, $f_{P,Q} = \frac{P^2}{2x_0\sin^2(Q\pi/P)}$ describes the degeneracy of the mode sequence, $A_{M,M'} = A_{m_0+MP,m_0+M'P}$ depicts the coupling matrix element produced by the tapered fiber, and $\gamma$ describes the coupling efficiency of the recombination states.

Figure 3 demonstrates the mode field intensity distribution plots calculated based on Eq. (8). The eigenvalue values $x_0$ ranging from 571 to 578 are calculated, according to the experimental conditions with the thin film thickness of 0.7 μm, microresonator diameter of 83.44 μm, and wavelength range from 970 nm to 980 nm. The triangle mode, dual-localization triangle mode and seven-star mode at wavelengths of 972.22 nm, 972.41 nm, and 973.09 nm are reproduced in Figs. 3(a), 3(b) and 3(c), respectively, when the distance between the tapered fiber and the microdisk center is tuned to 38.5 μm. When further increases the distance to 39.8 μm, the square mode, pentagon mode, hexagonal mode, heptagon mode and octagon mode are produced at wavelengths of 972.99 nm, 974.55 nm, 975.50 nm, 978.98 nm, and 979.95 nm, as depicted in Figs. 3(d)-3(h), respectively, indicating an excellent consistency with the experimental results.

To further illustrate the evolution characteristics of the degenerate polygon mode family $\{x_{p,m}^{(P,Q;M)}\}$ under the perturbation of the tapered fiber, here we take the triangle mode family $\{x_{63,289}^{(3,1;M)}\}$ as an example. Figure 4(a) gives the expectation values of the cosine of the chord angle Θ at different perturbation strengths. In the case of no perturbation, the expected cosines of chord angle are equally spaced around the classical orbit, and the cosines of chord angle is expected to be further away from the classical orbit as the number $M$ increases in absolute value. With the increase of perturbation strength, the expected cosines of chord angle are close to the cosines of the classical periodic orbits. The eigenvalue difference between mode with number $M$ and the center mode $x_{63,289}^{(3,1;0)}$ is illustrated in Fig. 4(b). In the case of no perturbation, the eigenvalues are inhomogeneousely distributed and the energy level interval increases as the increase of $M^2$. When further increases the perturbation strengths, the energy level interval of the modes increases gradually and tends to be uniform. Figures 4(c)-4(f) give typical phase space plots calculated with perturbation strengths of 0, 0.02, 0.1, and 0.3, respectively. The mode sequence $\{x_{63,289}^{(3,1;M)}\}$, originally distributed fully on the parabolic curve as shown in Fig. 4(c), gradually evolves into the linear distribution superimposed on the parabolic curve with the increasing perturbation strength as shown in Figs. 4(d)-4(f). Interestingly, the information lost in phase space of the mode is transferred to the spatial distribution. As shown in Figs. 4(h)-4(j), the orbits of the ground state (0th) and first excited state (1st) of the triangle become shaper with the increasing perturbation strength, forming the polygon mode observed in the experiment.

To conclude, the controllable trimming and clustering of high-Q WGMs in microdisk resonators provides a new control knob, which should have implication for the majority of research and application of WGM-based nonlinear photonics [22,23].

The work is supported by the National Key R&D Program of China (Grant No.2019YFA0705000), the National Natural Science Foundation of China (Grants No.62122079, 12192251, 11734009, 12134001, 11874375, 11874154, 12104159, 12174113, 11933005), Shanghai Municipal Science and Technology Major Project (Grant No.2019SHZDZX01), Science and Technology Commission of Shanghai Municipality (No.21DZ1101500), Innovation Program for Quantum Science and Technology (No. 2021ZD0301403) and the Youth Innovation Promotion Association of Chinese Academy of Sciences (Grant No. 2020249).

**Captions of figures:**

Fig. 1 (Color online) Experimental setup for investigating the polygon and star modes, and the observed mode patterns. (a) Experimental setup for investigating the polygon and star modes, the observed mode patterns of (b) triangle mode, (c) dual-localization triangle mode, (d) star mode, (e) square mode, (f) pentagon mode, (g) hexagonal mode, (h) heptagon mode, (i) octagon mode.

Fig. 2 (Color online) Cosine of the chord angle of the eigenmodes at different wavelengths. The black dots and colored dotted line represent the quasi-steady-state solution and the approximate result given by Eq. 5, respectively.

Fig. 3 (Color online) Mode field intensity distribution plots calculated by experimental conditions. The mode field intensity distributions of (a) triangle mode, (b) dual-localization triangle mode, (c) seven-star mode, (d) square mode, (e) pentagon mode, (f) hexagonal mode, (g) heptagon mode, (h) octagon mode.

Fig. 4 (Color online) Evolution characteristics of triangle mode with perturbation. (a) The expectation values of the cosine of the angle $\Theta$ and (b) the eigenvalue differences of modes at different perturbation strengths. (c)-(f) Phase space plots and (g)-(h) mode field distribution plots at different perturbation strengths. α: perturbation strength.

**Fig. 1**

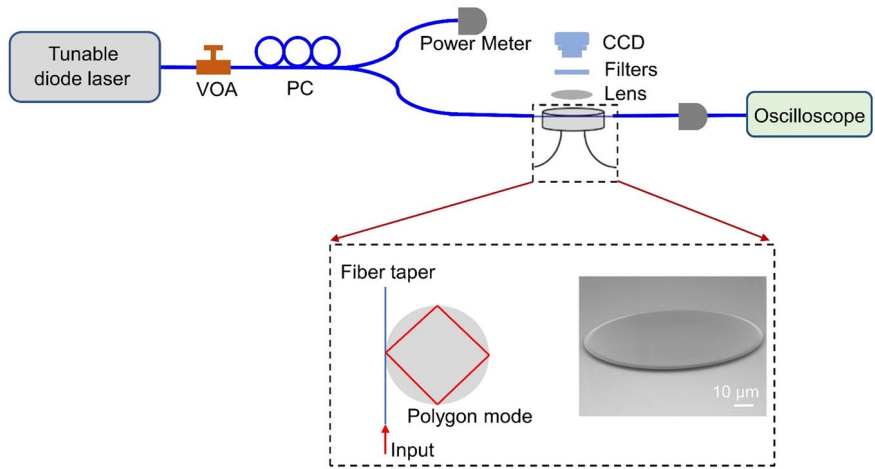

a

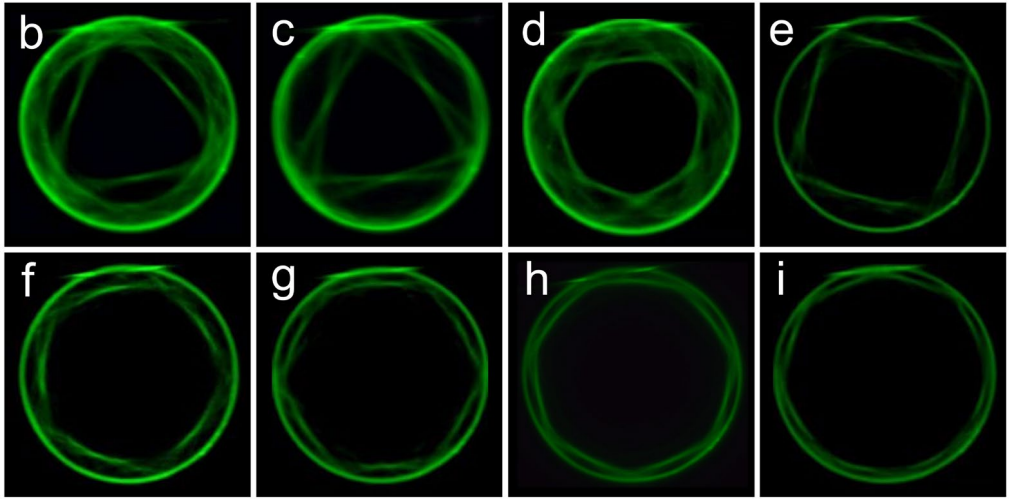

**Fig. 2**

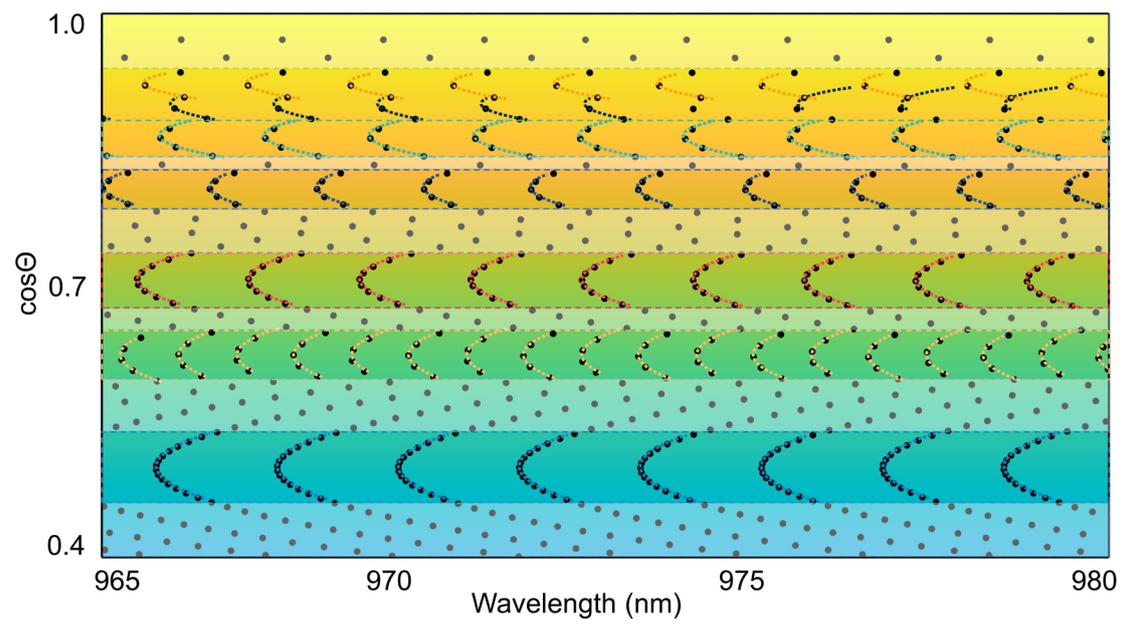

**Fig. 3**

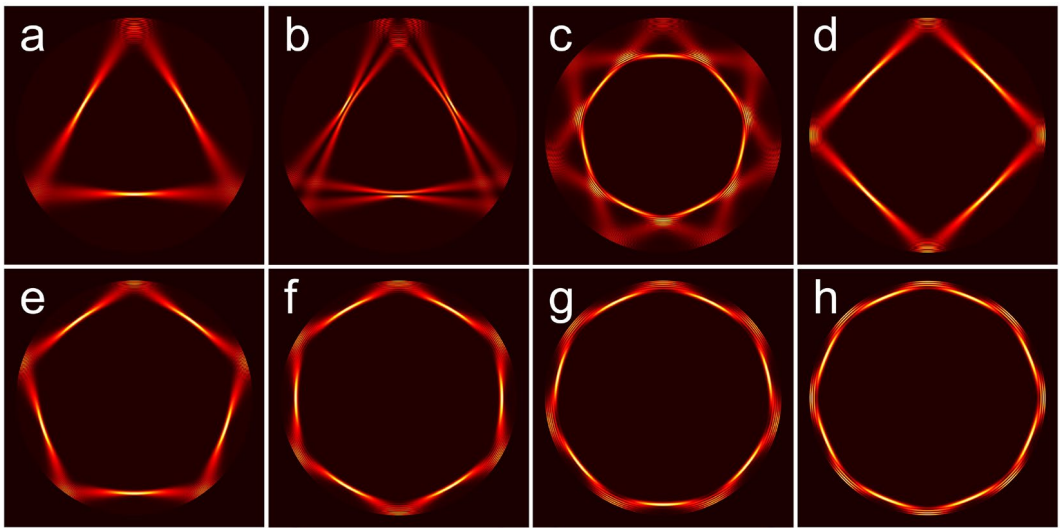

**Fig. 4**

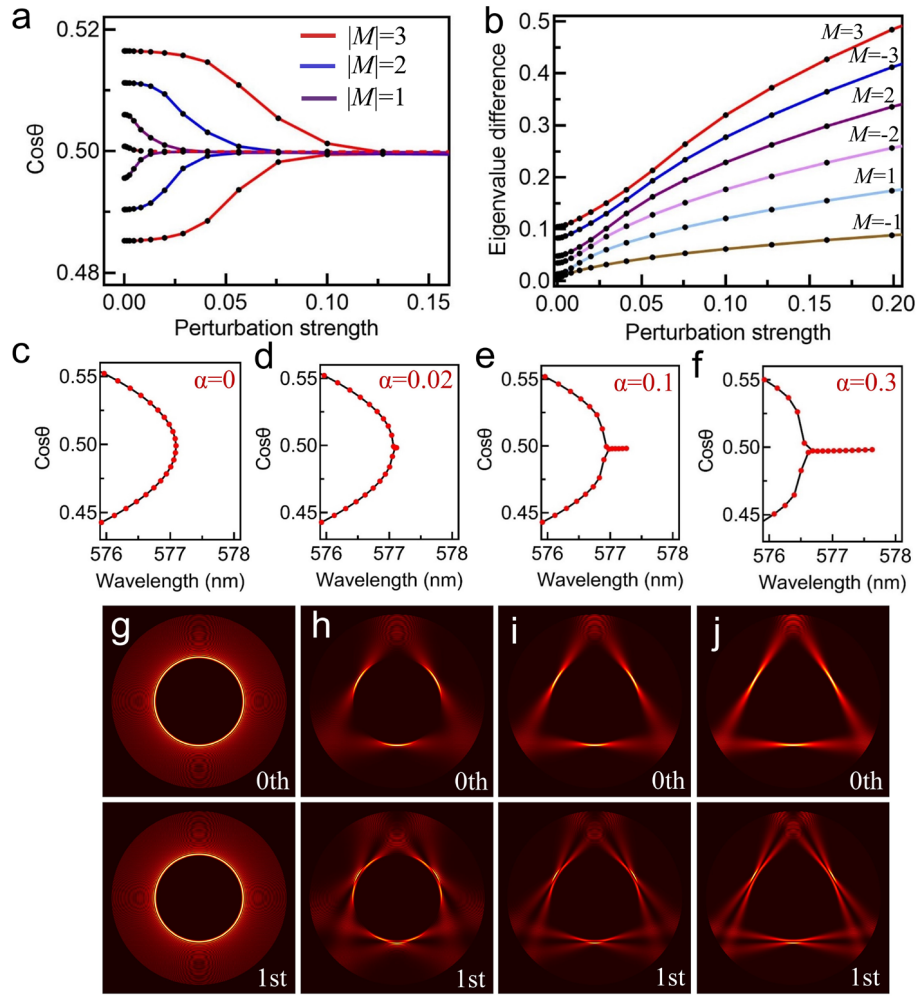